\documentclass{eas}
\usepackage{graphicx}  
\usepackage{dcolumn}   
\usepackage{bm}        
\usepackage[english]{babel}
\usepackage{amsfonts,amsmath,amssymb,mathrsfs}

\newcommand{\eq}{\begin{equation}}
\newcommand{\eeq}{\end{equation}}
\newcommand{\be}{\begin{equation}}
\newcommand{\ee}{\end{equation}}
\newcommand{\bea}{\begin{eqnarray}}
\newcommand{\nn}{\nonumber}
\newcommand{\eea}{\end{eqnarray}}
\newcommand{\cqg}{Class.\ Quant.\ Grav.\ }
\newcommand{\prd}{Phys.\ Rev.\ D\ }

\begin{document}


\title{Polytropic spheres in Palatini $f(R)$ gravity}
\runningtitle{Barausse \etal~: Polytropic spheres in Palatini $f(R)$ gravity}

\author{Enrico Barausse}\address{SISSA, International School for Advanced Studies, 
Via Beirut 2-4, 34014 Trieste, Italy and INFN, Sezione di Trieste}
\author{Thomas P.~Sotiriou}\sameaddress{1}\secondaddress{Department of Physics, University of Maryland, College Park, MD 20742-4111, USA}
\author{John C.~Miller}\sameaddress{1} \secondaddress{Department of Physics (Astrophysics), University of Oxford, Oxford, England}

\begin{abstract}
 We examine static spherically symmetric polytropic spheres in
Palatini $f(R)$ gravity and show that no regular solutions to the
field equations exist for physically relevant cases such as a
monatomic isentropic gas or a degenerate electron gas, thus casting
doubt on the validity of Palatini $f(R)$ gravity as an alternative to
General Relativity.
 \end{abstract}
\maketitle

\section{Introduction}

The quest for theories of gravity which can serve as alternatives to
General Relativity (GR) has become more intense due to recent feedback
from cosmology. The energy density of the universe appears to be
currently dominated by a cosmological constant, or by an unknown form
of energy (dark energy) that is mimicking the behaviour of a
cosmological constant~(\cite{carroll} 2001). The problems connected
with the inclusion of such a constant in Einstein's equations have
triggered research and the proposal of alternatives, one of which
could be to modify GR. One of the alternative theories which has been
considered is $f(R)$ gravity in the Palatini formalism. This formalism
consists of considering the metric and the connection as two
independent degrees of freedom and thus taking independent variations
of the action with respect to each of them in order to derive the field
equations, as opposed to the usual metric approach where the
connection is \textit{assumed} to be given by the Levi-Civita
connection of the metric and the action is varied only with respect to
the metric.  When applied to the usual Einstein-Hilbert action, this
procedure gives exactly Einstein's equations and the Levi-Civita 
formula for the connection, but when applied to a
generic action, the Palatini and metric variational approaches give
different results.  We consider here an action given
by $S=\int d^4 x\sqrt{-g}f(R)/16\,\pi+S_M(g_{\mu\nu},\psi)$ (with 
units in which $c = G = 1$)
where $R=g^{\mu\nu}R_{\mu\nu}$, $g$ is the determinant of the metric
$g_{\mu\nu}$, $S_M$ is the matter action and $\psi$ collectively
denotes the matter fields. We show that for generic choices of $f(R)$
there are natural matter configurations (such as spherical systems
composed of an isentropic monatomic gas or a nonrelativistic
degenerate electron gas) for which \textit{no} regular solution of the
field equations can be found if one adopts the Palatini variational
approach, apart from in the special case of GR, casting doubt on
whether Palatini $f(R)$ gravity can be considered as a viable
alternative to GR.

\section{A no-go theorem for polytropic spheres: the physics behind the proof}

To understand in which situations Palatini $f(R)$ gravity behaves
differently from GR, let us write the field equations using just
quantities built with the Levi-Civita connection of the metric (we
denote these with a ``tilde'', ) and \textit{not} with the
\textit{independent} connection. (As already mentioned, these expressions for
the connection are in general different in Palatini $f(R)$ gravity.) 
One then gets~(\cite{sot1} 2006b)
 \begin{align}
 \label{eq:field}
 \widetilde{G}_{\mu \nu} \!&=\! \frac{8\pi}{F}T_{\mu \nu}- \frac{1}{2}g_{\mu \nu}\! 
                         \left(\!R - \frac{f}{F} \right)\! +\! \frac{1}{F} \left(
 			\widetilde{\nabla}_{\mu} \widetilde{\nabla}_{\nu}
 			\!- g_{\mu \nu} \widetilde{\Box}
 		\right)\! F-\nn\\
 & \quad- \frac{3}{2}\frac{1}{F^2} \left(
 			(\widetilde{\nabla}_{\mu}F)(\widetilde{\nabla}_{\nu}F)
 			- \frac{1}{2}g_{\mu \nu} (\widetilde{\nabla}F)^2
 		\right),
 \end{align}
where $\widetilde{\Box}\equiv
g^{\mu\nu}\widetilde{\nabla}_{\mu}\widetilde{\nabla}_{\nu}$,
$F(R)=\partial f/\partial R$ and $T_{\mu\nu}\equiv -2(-g)^{-1/2}\delta
S_{M}/\delta g^{\mu\nu}$ is the usual stress-energy tensor of the
matter. The first three terms of this equation essentially give GR
plus a cosmological constant, while deviations away from it are
introduced by the terms depending on the first and second derivatives
of the function $F(R)$. It is important to note that the 
Ricci scalar $R$ (built with the \textit{independent} connection) is
algebraically related to the trace of the stress energy tensor,
because from the trace of the field equation one gets
 \begin{equation}
\label{trace}
F(R)R - 2f(R) = 8\,\pi\, T\,,
\end{equation}
 which can be solved for $R$. In vacuum, this equation shows that $R$
settles to a constant value $R_0$, and from Eq.~\eqref{eq:field} one
can see that Palatini $f(R)$ gravity in vacuum reduces to GR plus a
cosmological constant $\Lambda=R_0/4$. In the presence of matter
described by a perfect fluid with a 1-parameter equation of state
(EOS) $p=p(\rho)$ ($p$ and $\rho$ being the pressure and the energy
density of the fluid), Eq.~\eqref{trace} shows instead that $R$ can be
expressed as a function of $T=3p-\rho$ and therefore as a function of
$\rho$ alone. As such, the first and second derivatives of $F$
appearing in Eq.~\eqref{eq:field} involve first and second
derivatives of $\rho$. It is then apparent that Palatini $f(R)$
gravity will introduce important deviations away from GR when the
density changes rapidly. An obvious example of where this happens is given
by neutron stars, where the density changes rapidly when going from
the core to the inner crust and from the inner crust to the outer
crust. In~\cite{mainpaper} (2007a) we have indeed studied static and
spherically symmetric neutron star models in Palatini $f(R)$ gravity,
and have found that the deviations away from GR can be very
important for forms of $f(R)$ such as those expected from cosmology.
Another place were the derivatives of the density become large is at the
surface of polytropic spheres. [We recall that a polytropic EOS has
the form $p=\kappa \rho_0^\Gamma$, with $\rho_0$ being the rest-mass
density and $\kappa$ and $\Gamma$ ($>1$) being two constants; this can 
be written in the
equivalent form $\rho=(p/\kappa)^{1/\Gamma}+p/(\Gamma-1)$.] To see
this, let us note that from the field equations~\eqref{eq:field} it
follows that the stress energy tensor is conserved under covariant 
differentiation using the
Levi-Civita connections of the metric, \textit{i.e.}
$\tilde{\nabla}_\nu T^{\mu\nu}=0$.\footnote{Alternatively this can be
derived as in GR from the diffeomorphism invariance of the matter
action, see for instance~\cite{defelice} (1990), section 6.3.} Inserting a perfect fluid stress energy tensor and a static
spherically symmetric ansatz for the metric into this equation, one
gets the usual Euler equation $p'=-A'(p+\rho)/2$, where we denote
radial derivatives with a ``prime'' and $g_{tt}\equiv \exp[A(r)]$.
Using now the polytropic EOS in the Euler equation: near to the 
surface (defined as where $p=0$) of a polytropic sphere, one gets 
$p'\propto
p^{1/\Gamma}$, hence $p\propto(r_{\rm out}-r)^{\Gamma/{\Gamma-1}}$
($r_{\rm out}$ being the radius of the sphere) and, from the
polytropic EOS, $\rho\propto(r_{\rm out}-r)^{1/{\Gamma-1}}$. It is now
trivial to check that the second radial derivative of the density, 
$\rho''$, diverges for $\Gamma>3/2$, and one thefore expects major
differences between Palatini $f(R)$ gravity and GR in this case. In
fact, it is possible to show~(\cite{mainpaper} 2007a) that the
corrections coming from the terms depending on the derivatives of $F$ in
Eq.~\eqref{eq:field} become so important that they make the curvature
invariants $\widetilde{R}$ and
$\widetilde{R}^{\mu\nu\sigma\lambda}\widetilde{R}_{\mu\nu\sigma\lambda}$
diverge. As such, there exist \textit{no} regular solutions to the
field equations of Palatini $f(R)$ gravity for polytropic spheres with
$\Gamma>3/2$, apart from in the special case of GR (which 
does not have this kind of problem because only the density, and not
its derivatives, enters the field equations). Physically, this means
that the tidal forces diverge at the surface of such objects, although
the density is exactly zero there. 

It was recently argued~(\cite{finns} 2007) that the singularity which
we found would not cast doubt on the viability of Palatini $f(R)$
gravity because of the idealized nature of the polytropic EOS and
because the lengthscale on which the tidal forces diverge due to the
singularity would be shorter than the lengthscale on which the fluid
approximation is valid, \textit{i.e.} the mean free path (MFP). While
it is true that the polytropic EOS may be too idealized to describe
the outer layers of an astrophysical star, we note that $\Gamma=5/3$,
corresponding to an isentropic monatomic gas or a degenerate
non-relativistic particle gas, falls within the range not giving a
regular solution. These are perfectly physical configurations which
should be describable by a viable theory of gravity without resorting
to further considerations of the microphysics.  Alternatively, one 
should accept that the theory is at best \textit{incomplete} because 
of being unable to describe
configurations which are well-described even by Newtonian gravity.
Note that this means in particular that Palatini $f(R)$ gravity does
not reproduce the Newtonian limit! 

About the lengthscale on which the tidal forces diverge: in a
forthcoming paper~(\cite{nextpaper} 2007b) we will show that while
this is smaller than the MFP in the particular case considered
in~\cite{finns} (2007) (\textit{i.e.} a neutron star with
$f(R)=R-\mu^4/R$, where $\mu^2\sim \Lambda$, with $\Lambda$ being the
cosmological constant needed to explain the accelerated
expansion of the universe), this is a very special situation. While
$f(R)=R-\mu^4/R$ can explain cosmological data
without Dark
Energy, there is no first principle from which to derive this
functional form, and in order to justify it one has to invoke
arguments based on the series expansion of the
unknown $f(R)$ coming from a consistent high energy theory. As such,
there is no reason to exclude the presence, in the function $f(R)$, of
terms quadratic or cubic in $R$, and the constraints on these
terms coming from solar system tests are rather loose~(\cite{tomo}
2006a). In~\cite{nextpaper} (2007b), we will show that if
$f(R)=R-\mu^4/R+\varepsilon R^2$ (with $\varepsilon$ even several orders of
magnitude lower than the solar system constraint), the scale on which
the tidal forces diverge is \textit{much} larger than the MFP, even
for neutron stars. [This was expected: we have already shown
in~\cite{mainpaper} (2007a) how the effect of such a tiny
$\varepsilon$ can be important in neutron star interiors]. However,
even if one cancels \textit{by hand} all of the quadratic and cubic terms
from the function $f(R)$, so that $f(R)=R-\mu^4/R$, the claim
of~\cite{finns} (2007) still does not apply for sufficiently diffuse
systems, where the scale on which the tidal forces diverge is 
\textit{much} larger than the MFP.

\section{Conclusion}
The problems discussed here arise due to the dependence of the metric
on higher order derivatives of the matter fields, and we can expect
that any theory having a representation in which the field equations
include second derivatives of the metric and higher than first
derivatives of the matter fields will face similar problems. The same
should be expected for theories which include fields
other than the metric for describing the gravitational interaction
(\textit{e.g.}~scalar fields) which are algebraically related to
matter rather than dynamically coupled. Indeed, one can 
solve the field equations for the extra field and insert the solution
into the equation for the metric, inducing a dependence of the
metric on higher derivatives of the matter fields. An example is a
scalar-tensor theory with Brans-Dicke parameter $\omega=-3/2$, which
is anyway an equivalent representation of Palatini $f(R)$
gravity~(\cite{sot1} 2006b).  As such, our
results cast doubt on the viability of theories including higher order
derivatives of the matter fields in one of their representations, such
as generic Palatini $f(R)$ gravity or $\omega=-3/2$ scalar-tensor 
theory.

\enlargethispage{\baselineskip}

\end{document}